\documentclass{article}
\usepackage{spconf}

\usepackage{amsmath,graphicx}

\usepackage{hhline}
\usepackage{siunitx}

\usepackage{pict2e}

\usepackage{ctable}

\usepackage{balance}

\title{Emotion Classification: How Does an Automated System Compare to Naive Human Coders?}

\begin{document}
\ninept 

\name{\em{Sefik Emre Eskimez$^{\star}$} \qquad Kenneth Imade$^{\star}$ \qquad Na Yang$^{\star}$ \qquad Melissa Sturge-Apple$^{\dagger}$ \\
	\em{Zhiyao Duan$^{\star}$} \qquad Wendi Heinzelman$^{\star}$}

\address{$^{\star}$Dept. of Electrical and Computer Engineering, \\
	$^{\dagger}$Dept. of Psychology, University of Rochester, Rochester, NY
	}

\maketitle

\begin{abstract}
	The fact that emotions play a vital role in social interactions, along with the demand for novel human-computer interaction applications, have led to the development of a number of automatic emotion classification systems. However, it is still debatable whether the performance of such systems can compare with human coders. To address this issue, in this study, we present a comprehensive comparison in a speech-based emotion classification task between 138 Amazon Mechanical Turk workers (Turkers) and a state-of-the-art automatic computer system. The comparison includes classifying speech utterances into six emotions (happy, neutral, sad, anger, disgust and fear), into three arousal classes (active, passive, and neutral), and into three valence classes (positive, negative, and neutral). The results show that the computer system outperforms the naive Turkers in almost all cases. Furthermore, the computer system can increase the classification accuracy by rejecting to classify utterances for which it is not confident, while the Turkers do not show a significantly higher classification accuracy on their confident utterances versus unconfident ones. 
		
\end{abstract}

\begin{keywords}
	Voice-based Emotion Classification, Amazon Mechanical Turk, Human vs. Machine
\end{keywords}

\section{Introduction} \label{Introduction}

	\let\thefootnote\relax\footnotetext{\copyright\ Copyright 2016 IEEE. Published in the IEEE 2016 International Conference on Acoustics, Speech, and Signal Processing (ICASSP 2016), scheduled for 20-15 March 2016 in Shanghai, China. Personal use of this material is permitted. However, permission to reprint/republish this material for advertising or promotional purposes or for creating new collective works for resale or redistribution to servers or lists, or to reuse any copyrighted component of this work in other works, must be obtained from the IEEE. Contact: Manager, Copyrights and Permissions / IEEE Service Center / 445 Hoes Lane / P.O. Box 1331 / Piscataway, NJ 08855-1331, USA. Telephone: + Intl. 908-562-3966.}
Emotion classification is a fundamental task for humans in order to interpret social interactions. Although emotions are expressed at various levels (e.g., behavioral, physiological), vocal and verbal communication of emotions  is a central domain of communication research \cite{scherer1986vocal}. Classification accuracy is essential in order to be ensured of the validity and reliability of emotional constructs used in psychological research.
Given the importance of accurately classifying emotions to understanding human interactions, many researchers have developed automatic emotion classification computer systems.  There are a number of modalities that can be used to determine one's emotions, including facial expression, body movement, physiological measures such as galvanic skin response, and voice.  While automatic emotion classification systems have been developed that use all of these modalities, individually and in concert \cite{Busso:2004:AER:1027933.1027968,huisman2013lemtool,ozkul2012multimodal,wu2013two}, several systems have focused on classifying emotions using speech features in particular \cite{yang2012speech, bitouk2010class,rachuri2010emotionsense,sethu2008empirical}.  There are a number of reasons for this, including the fact that speech is relatively easy to capture and is less intrusive than other methods for capturing emotional state.  While these speech-based automatic emotion classification systems all provide reasonable accuracy in their classification results, it is not known how well these systems, which in many applications would replace a human's classification of the emotion, compare to a naive human coder performing the same emotion classification task. \par

In this work, we compare how well an automated computer system can perform at the task of emotion classification from speech samples compared with naive human coders.  In particular, we asked Amazon Mechanical Turk workers (Turkers) to listen to speech samples from the LDC dataset of emotions \cite{LDC} and classify them in three ways: 1) determine whether the conveyed emotion was active, passive or neutral; 2) determine whether the conveyed emotion was positive, negative or neutral; and 3) determine which of six emotions (happy, neutral, sad, anger, disgust, fear) was being conveyed.  We also asked the Turkers how confident they were in their classification.  We compared the Turkers' accuracy with that achieved by a speech-based automated emotion classification system \cite{yang2012speech}, using a leave-one-subject-out (LOSO) approach for evaluating the system. \par
Our results show that the automated system has a higher emotion classification accuracy compared with the Turkers' accuracy averaged over all six emotions, with the automated system able to achieve close to 72\% accuracy compared with the Turkers' accuracy of only about 60\%.  Additionally, while the automated system can achieve even better accuracy by rejecting samples when its confidence in the classification is low, the Turkers' results for the samples in which they were confident about their classification did not show any significant improvement compared with the accuracy of all their responses. These results suggest that an automated speech-based emotion classification system can potentially replace humans in scenarios where humans cannot be easily trained.
\section{Related Work} \label{Related Work}
To date, only a few studies have been conducted to compare the performance of automatic systems with that of humans for emotion classification. Some of these studies use visual facial expressions to determine emotion \cite{holkamp2014comparison,janssen2013machines, susskind2007human}, but these are out of the scope of this study, which focuses on comparing human and machine performance for emotion classification based on speech. \par
For the existing studies on human emotion classification from speech, the number of human subjects used is relatively small. In addition, whether the human subjects were trained for the specific emotion classification task or not is not always specified. 
In \cite{shaukat2010emotional}, Shaukat et al. compared a psychology-inspired automatic system that utilizes a hierarchical categorization model based on multiple SVMs with humans' ability to classify emotions from speech on two databases, the Serbian Emotional Speech Corpus (GEES) and the Danish Emotional Speech Corpus (DES). For the experiments with humans, there were 30 subjects for the GEES, and 20 subjects for the DES. Results showed that the automatic system slightly underperformed humans for both databases. \par
In \cite{esparza2012automatic}, Esparza et al. employed a multi-class SVM system to classify speech emotions, and compared its performance with humans on two German databases, the ``corpus of spoken words for studies of auditory speech and emotional prosody processing'' (WaSeP), and the Berlin Database of Emotional Speech (EmoDB). The WaSeP corpus was evaluated by 74 native German speakers with an accuracy of 78.5\%, and the EmoDB corpus was evaluated by 20 native German speakers with an accuracy of 84.0\%. Computer system accuracies were 84.0\% and 77\% for the WaSeP and EmoDB databases, respectively. In this case the results (whether humans or the automated system perform better) were mixed.
A final study considered a Hungarian database evaluated by both humans and an automated emotion classification system that utilized Hidden Markov Models (HMMs) \cite{toth2008speech}. The evaluation was performed by 13 subjects, where the subjects never heard the same speaker successively. The evaluation included utterances that contained emotion as well as neutral utterances. The authors evaluated the 4 best emotional categories for the computer system with average accuracy around 85\%, and they evaluated the 8 best emotion categories for the human subjects, with average accuracy of 58.3\%. The results showed that the humans provided better evaluations for the sad and disgust emotion categories, while the computer system provided better evaluations for the surprised and nervous emotion categories. \par
In this paper, we conducted a large scale comparison between a
state-of-the-art speech-based emotion classification system with the
performance of 138 human subjects classifying 7270 audio samples. These human subjects were recruited using the Amazon Mechanical Turk service. Compared to existing studies, our experiment used more human subjects
with much higher diversity both demographically and geographically.
In addition, these human subjects were not trained on the
dataset used in the experiment. 
\section{LDC Dataset} \label{LDC Dataset}
In this study, we use the LDC dataset, a collection that includes speech samples with 14 distinct emotion categories recorded by professional actors, 3 male and 4 female, reading semantically neutral utterances such as dates and times \cite{LDC}. Note that using semantically neutral utterances is a common practice in speech-based human emotion classification studies \cite{esparza2012automatic, shaukat2010emotional,toth2008speech}. The length of the utterances varies between one and two seconds. In our study, we used a total of 727 utterances that contained the emotions happy (136), neutral (67), sad (157), anger (136), disgust (108), and fear (123).  Each emotion was also labeled as active (happy, anger, fear), passive (sad, disgust) and neutral as well as positive (happy), negative (sad, anger, disgust, fear) and neutral.	
\section{Automated Emotion Classification System} \label{EmoSystemSection}
There are a number of systems that automatically classify emotions from speech \cite{bitouk2010class,  rachuri2010emotionsense, schuller2003hidden, sethu2008empirical, yun2012loss}.
In this paper, we use the one described in \cite{yang2012speech}, as it has been shown to achieve similar or better classification accuracy than several other state-of-the-art systems \cite{bitouk2010class, rachuri2010emotionsense, sethu2008empirical} and it has the added advantage that it can reject samples as unclassified if it is not a confident classification. The rejection mechanism is useful in scenarios where classification is not required on all samples and the cost of an incorrect classification is high; hence, it is better to simply not classify some samples in order to achieve a much higher classification accuracy on all classified samples. Here, we briefly overview this emotion classification system. \par
In this system, speech utterances are divided into 60 ms frames with a hop size of 10 ms. For all voiced frames (frames that contain voiced speech), several features are calculated, including: fundamental frequency ($F_0$), energy, frequency and bandwidth of the first four formants, and 12 mel-frequency cepstral coefficients (MFCCs), zero crossing rate, spectral rolloff, brightness, centroid, spread, skewness, kurtosis, flatness, entropy, roughness, irregularity and the derivative of all features \cite{Signal2006Process}. Five statistics of these features (mean, standard deviation, min, max, and range) are then calculated over all speech frames to obtain utterance-level features. Additionally, speaking rate is calculated for each utterance. This provides a total of 331 features for each utterance. \par
A classification system with 6 one-against-all (OAA) support vector machines (SVM), one for each emotion, with radial basis function (RBF) kernels, is then trained using the features extracted from training data together with their ground-truth emotion labels. This system is then able to classify new unseen utterances.  For an unseen utterance, each OAA classifier outputs a confidence value, indicating the classifier's confidence that the utterance conveys that particular emotion. The confidence values of all 6 classifiers are compared, and the final emotion label of the utterance is determined by the classifier with the highest confidence. \par
In many scenarios, a classification does not have to be made for every utterance, yet when a classification is made, the cost of an incorrect classification is high. To deal with these scenarios, the system is also equipped to perform thresholding fusion, as per the approach in \cite{vapnik1998statistical}. If the highest confidence value is below a threshold, the system rejects the sample. Only if the confidence value is above a threshold will the system provide a label for the utterance. \par
The system also employs speaker normalization, training set over-sampling, and feature selection to enhance the classification performance \cite{yang2012speech}. Speaker normalization (z-score normalization \cite{farrus2007histogram}) is used to normalize the distribution of the features of each speaker. This is to cope with the problem that different speakers may have distinct speech characteristics such as loudness. Training set over-sampling is used to overcome the problem of having an unbalanced training. SMOTE \cite{chawla2002smote} over-sampling method is used, where synthetically created samples are added to the training set to balance the training data set. Feature selection is employed to select the most effective features from the 331 features for the classification. While in prior work \cite{yang2012speech}, Mutual Information (MI) was used, here we use an SVM Recursive Feature Elimination \cite{guyon2002SVMRFE} method instead, as we found that this approach can provide overall better performance in terms of classification accuracy using a subset of the features. \par
Similar OAA-SVM classification systems were trained for active-passive-neutral (APN) and positive-negative-neutral (PNN). These systems also use the thresholding fusion mechanism to reject utterances for which they are not confident enough, in order to improve the classification accuracy of those utterances that are ultimately classified. \par	
To evaluate these systems, we conduct leave-one-subject-out (LOSO) tests, where the OAA binary classifiers are trained using speech utterances from all but one of the speakers in the dataset, and then tested using the one speaker left out of the training phase. In this way, we can determine the performance of the system when it has not been trained on the individual speaker, as would be the case for a number of applications where the system can be trained on the class of speakers it will encounter but it cannot be trained using samples from the target person.  The results represent the average over all 7 LOSO cases using the 7 speakers in the LDC dataset.
\begin{figure}[t!]
	\centering
	\includegraphics[scale=0.4]{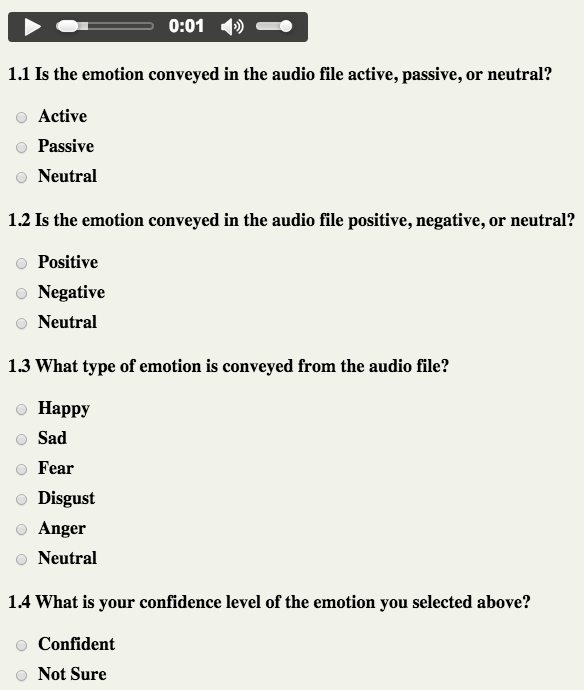}	
	\caption{Questions shown to Turkers.}
	\label{fig:MTurk_Page_2}
\end{figure}

\section{Amazon's Mechanical Turk Setup} \label{AMTSection}
For the MTurk experiment, each Turker was provided with initial instructions about the task. These instructions included a sample of each of the different emotion categories to provide some minimal training of the Turkers. After reviewing the instructions, the Turkers were presented with a random selection of 10 to 100 audio samples to classify. After listening to each audio sample, as shown in Figure \ref{fig:MTurk_Page_2}, Turkers were asked if the emotion conveyed in the sample was 1) active, passive or neutral, 2) positive, negative or neutral, and 3) one of the six emotions.  Additionally, the Turkers were asked to rate their confidence in their labeling of the emotion.  Finally, they were asked to transcribe what they heard in the audio file in order to ensure that the Turker actually listened to the audio sample. After completing the classification of all audio samples, the Turkers were asked to provide demographic information, including gender, age and race. Note that only Turkers whose native language is English are requested for this  task. Once the MTurk task was completed and approved by us, the Turkers were paid \$0.50 for each group of 10 speech samples they classified, in exchange for their time. \par
Table \ref{table:WCount} shows the number of samples classified by the Turkers, broken down by gender and age. There were 138 unique Turkers that classified 7,270 audio samples, with individual Turkers classifying between 10 and 100 audio samples.

\section{Evaluation} \label{Evaluation}
The goal in our experiments is to compare the performance of the automatic emotion classification system described in Section \ref{EmoSystemSection} with the performance of naive human coders, the Turkers from our MTurk experiment described in Section \ref{AMTSection}.  In order to provide a fair comparison, we present results for the leave-one-subject-out (LOSO) case for the automatic emotion classification system.  In this case, the training set does not contain any samples from a particular speaker who is used in the test set. \par 
In this section, we compare three different ways of classifying the utterances: 1) classifying the utterances to one of 6 different emotions; 2) classifying the utterances as active, passive or neutral; and 3) classifying the  utterances as positive, negative or neutral.  For each case, we compare the results of the automatic system with the Turkers when classifying all samples as well as when classifying only those samples for which they are confident.  Additionally, we provide data showing the performance of the male and female Turkers, and for the performance for female and male utterances separately.

\subsection{Classifying the Utterances: 6 Emotions}
The first task is to classify the samples into the six emotion categories mentioned in Section \ref{LDC Dataset}. Table \ref{table:Emotion} shows the accuracy values for the computer system and the Turkers for this task. Overall, the average accuracy for the Turkers is 60.4\%, which is 12.5\% worse than the automatic emotion classification system, which provides an accuracy of 72.9\%. As shown in this table, the Female Turkers performed slightly better than the Male Turkers (1.1\% improvement).\par		
Also shown in Table \ref{table:Emotion} are the different accuracy values for the computer system and the Turkers when considering only the female or male utterances.  It is interesting to note that the computer system performs slightly better (1.2\%), while the Turkers perform significantly better (10.8\%) for the female utterances. \par	
We compare the accuracy values for the samples where the Turkers were confident about their classification with the accuracy values when the automatic emotion classification system is confident (here we use two different thresholds such that either 50\% or 80\% of the samples are classified, with all others being rejected). If we compare the Turkers' accuracy in classifying the emotions when all samples are classified with the accuracy when only those samples for which they were confident in their classification are considered, we see very little difference in the accuracy values (60.4\% vs. 60.6\%). This tells us that humans are not able to accurately estimate their performance and reliability on the emotion classification task.  On the other hand, if we look at the automatic emotion classification system results, we see that when the computer rejects as unclassified the samples for which the confidence values from the OAA SVM are low, the accuracy of those samples that are classified increases from 72.9\% to 77.7\% when 80\% of the samples are classified and to 85.4\% when 50\% of the samples are classified.  Hence, we see that one clear advantage of an automatic emotion classification system over human coders is this ability to improve classification accuracy by rejecting to classify some samples.  In applications where not all samples must be classified and the cost of mis-classification is high, this can be a valuable means to increase emotion classification accuracy. \par
\begin{table}[t!]
	\centering
	\caption{Number of samples classified by Turkers.}
	\label{table:WCount}
	\resizebox{\columnwidth}{!} {
		\begin{tabular}{|c|c|c|c|c|c|c|c|c|}
			\hline
			\multicolumn{9}{|c|}{{\bf All (7270)}}                                                                                       \\ \hline
			
			\multicolumn{4}{|c|}{{\bf Female (2850)}}                    & \multicolumn{4}{c|}{{\bf Male (4350)}}                       & {\bf NA (70)}    \\ \hline
			\multicolumn{4}{|c|}{{\bf Ages}}                      & \multicolumn{4}{c|}{{\bf Ages}}                       & {\bf Ages}  \\ \hline
			{\bf 18-29} & {\bf 30-39} & {\bf 40-49} & {\bf 50-59} & {\bf 18-29} & {\bf 30-39} & {\bf 40-49} & {\bf 50-59} & {\bf 18-29} \\ \hline
			1300         & 630          & 620          & 300          & 2610         & 940          & 550          & 250          & 70           \\ \hline
		\end{tabular}}
	\end{table}
\begin{table}[t!]
	\centering
	\caption{Accuracy values (\%) for six emotions.}
	\label{table:Emotion}
	\resizebox{\columnwidth}{!} {
	\begin{tabular}{|l|c|c|c|c|c|c|c|}
		\hline
		&               & \multicolumn{2}{c|}{{\bf \begin{tabular}[c]{@{}c@{}}Speaker \\ Gender\end{tabular}}} & \multicolumn{4}{c|}{{\bf \begin{tabular}[c]{@{}c@{}}Classification \\ Confidence Level\end{tabular}}}                                                                                                                                                               \\ \hline
		{\bf Accuracy}  & {\bf Overall} & {\bf Female}                               & {\bf Male}                              & {\bf \begin{tabular}[c]{@{}c@{}}Confident\\ (80\%)\end{tabular}} & {\bf \begin{tabular}[c]{@{}c@{}}Confident\\ (50\%)\end{tabular}} & {\bf \begin{tabular}[c]{@{}c@{}}Unsure\\ (20\%)\end{tabular}} & {\bf \begin{tabular}[c]{@{}c@{}}Unsure\\ (50\%)\end{tabular}} \\ \hline
		Computer System & 72.9          & 73.2                                       & 72.0                                    & 77.7                                                             & 85.4                                                             & 61.2                                                          & 55.3                                                          \\ \hline
		All Turkers     & 60.4          & 64.9                                       & 54.1                                    & \multicolumn{2}{c|}{60.6 (80.5\% confident)}                                                                                        & \multicolumn{2}{c|}{59.6 (19.5\% unsure)}                                                                                     \\ \hline
		Female Turkers  & 61.2          & 64.4                                       & 57.1                                    & \multicolumn{2}{c|}{60.4 (78.4\% confident)}                                                                                        & \multicolumn{2}{c|}{62.9 (21.6\% unsure)}                                                                                     \\ \hline
		Male Turkers    & 60.1          & 65.4                                       & 52.5                                    & \multicolumn{2}{c|}{60.8 (82.0\% confident)}                                                                                        & \multicolumn{2}{c|}{57.9 (18.0\% unsure)}                                                                                     \\ \hline
	\end{tabular}
	}
\end{table}
\begin{table}[t!]
	\centering
	\caption{Confusion matrix for the automatic classification system (GT = ground truth).}
	\label{table:ConfusionOurSystemGA}
	\resizebox{\columnwidth}{!} {
		\begin{tabular}{|l|c|c|c|c|c|c|}
			\hline
			{\bf }             & {\bf Anger} & {\bf Disgust} & {\bf Fear} & {\bf Happy} & {\bf Neutral} & {\bf Sad} \\ \hline
			{\bf Anger (GT)}   & 92.9          & 0.0             & 2.4          & 2.5           & 0.0             & 2.2        \\ \hline
			{\bf Disgust (GT)} & 0.9           & 80.7            & 0.9          & 6.0           & 1.1             & 10.3     \\ \hline
			{\bf Fear (GT)}    & 4.3           & 0.0             & 85.2         & 8.9           & 0.0             & 1.6       \\ \hline
			{\bf Happy (GT)}   & 5.6           & 3.5             & 8.2          & 79.0          & 1.5             & 2.2       \\ \hline
			{\bf Neutral (GT)} & 0.0           & 4.2             & 0.0          & 2.4           & 86.3            & 7.2      \\ \hline
			{\bf Sad (GT)}     & 0.0           & 5.1             & 0.0          & 0.8           & 1.5             & 92.6      \\ \hline
		\end{tabular}
	}
\end{table}
		
The final set of numbers shows the accuracy of the utterances that are rejected by the automatic classification system or for which the Turkers were unsure of their classification.  From this data, we can see that there is not much difference in accuracy for the set where the Turkers were confident (60.6\%) and for the set where the Turkers were not confident (59.6\%).  Additionally, this data shows that when 20\% of the samples are rejected by the automatic classification system, the accuracy on those rejected samples is 55.3\%. Hence, some of the rejected samples (55.3\%) were actually correctly classified.  However, it is impossible to know which ones, and including this set of classifications makes the overall classification accuracy drop, and in some applications this is not a good trade-off to make. Nevertheless, it is interesting to see that the computer system's accuracy on the rejected samples is very close to that obtained even by confident Turkers, which further shows the superiority of the computer system over naive human coders on this dataset. \par
Confusion matrices for the automatic emotion classification system's classification and the Turkers' classification for the 6 emotions are shown in Tables \ref{table:ConfusionOurSystemGA} and \ref{table:ConfusionMTURKAll}, respectively. Note that in these tables, the rows are the ground truth (GT) labels, and they sum to 100\%. From these tables, we see that the automatic classification system is classifiying each emotion better than the Turkers.

\begin{table}[t!]
	\centering
	\caption{Confusion matrix for the Turkers (GT = ground truth).}
	\label{table:ConfusionMTURKAll}
	\resizebox{\columnwidth}{!} {
		\begin{tabular}{|l|c|c|c|c|c|c|}
			\hline
			{\bf }             & {\bf Anger} & {\bf Disgust} & {\bf Fear} & {\bf Happy} & {\bf Neutral} & {\bf Sad} \\ \hline
			{\bf Anger (GT)}   & 69.0        & 14.7          & 4.6        & 6.8         & 3.5           & 0.7       \\ \hline
			{\bf Disgust (GT)} & 7.8         & 32.5          & 9.4        & 6.8         & 28.0          & 15.0      \\ \hline
			{\bf Fear (GT)}    & 11.2        & 3.6           & 67.2       & 11.3        & 4.2           & 2.3       \\ \hline
			{\bf Happy (GT)}   & 3.3         & 6.3           & 8.0        & 54.7        & 22.9          & 4.3       \\ \hline
			{\bf Neutral (GT)} & 0.9         & 2.1           & 0.4        & 1.8         & 78.4          & 15.8      \\ \hline
			{\bf Sad (GT)}     & 0.5         & 3.7           & 5.6        & 0.3         & 25.2          & 64.3      \\ \hline
		\end{tabular}
	}
\end{table}
\begin{table}[t!]
	\centering
	\caption{Accuracy values (\%) for APN and PNN.}
	\label{table:APN}
	\resizebox{\columnwidth}{!} {
		\begin{tabular}{|l|c|c|c|c|c|}
			\hline
			& \multicolumn{3}{c|}{{\bf Samples}}                        & \multicolumn{2}{c|}{{\bf Classification Confidence}}  \\ 	\hline
			& {\bf All} &{\bf Female} &{\bf Male} &{\bf \begin{tabular}[c]{@{}c@{}}Confident\\ (80\%)\end{tabular}} &{\bf \begin{tabular}[c]{@{}c@{}}Unsure\\ (20\%)\end{tabular}} \\ \hline
			{\bf Computer (APN)} & 89.3        & 86.8           & 92.4         & 94.4                      & 73.1                   \\ \hline
			{\bf Turkers (APN)}         & 70.5        & 71.5           & 69.0         & 71.0                      & 67.9   \\	\hhline{|=|=|=|=|=|=|}   
			{\bf Computer (PNN)} & 82.9        & 82.9           & 82.4         & 88.0                      & 62.0                   \\ \hline
			{\bf Turkers (PNN)}         & 71.8        & 75.5           & 66.6         & 72.1                      & 70.7   \\	\hline                 
		\end{tabular}
	}
\end{table}
\subsection{Classifying into Active-Passive-Neutral }
Next, we explore the results when classifying the samples according to the three arousal categories: active, passive and neutral (APN). As can be seen in Table \ref{table:APN}, the Turkers achieved 70.5\% accuracy in their classification of the utterances into active, passive and neutral categories, while the computer system achieved 89.3\% accuracy. As for the 6 emotion classification task, the accuracy for the samples for which the Turkers are confident in their classification does not improve significantly compared with the accuracy for all the samples, while the computer system does have an increase in accuracy when only classifying samples for which it is confident in the response.
\subsection{Classifying into Positive-Negative-Neutral }
For the final classification task, we explore the results when classifying the samples according to the three valence categories: positive, negative and neutral (PNN). As can be seen in Table \ref{table:APN}, the Turkers achieved 71.8\% accuracy in their classification of the utterances into positive, negative and neutral categories, while the computer system achieved 82.9\% accuracy. Once again, the same conclusions hold for the confident utterances.	
\section{Discussions}
It is important to note that the expression and perception of emotion are very subjective. For the same utterance, different listeners may perceive different emotions, and all of them may be different from the emotion that the speaker intended to convey. Therefore, for an emotion classification task, obtaining the ground-truth emotion labels is not trivial. To obtain the ground-truth ``perceived'' emotion, one could ask some listeners to label the utterance, but these labels are ambiguous due to their subjective nature. Our results also show that different Turkers do sometimes disagree with each other. \par
Due to this difficulty in obtaining ground-truth emotion labels, in our study we used acted emotions. On the one hand, one may criticize that these utterances may not be ``natural''. On the other hand, however, the ground-truth labeling is not an issue. Each utterance is labeled to the emotion that the speaker wants to convey, hence the ground-truth labels are ``expressed'' emotions. Consequently, the classification errors that the Turkers made simply indicate the mismatches between the emotions that the speakers wanted to convey and the emotions that the Turkers perceived, i.e., the effectiveness of the emotion communication through these utterances. \par
Compared to the automated classification system, emotion communication between humans is apparently less effective according to the results in our study. One of the most important reasons, we argue, is due to the lack of training. The computer system was trained and tested on the same dataset. Although utterances from different speakers were used for training and testing, they did share some common characteristics such as the type of speech content and the recording environment. The Turkers, however, were only provided with 1 sample recording for each emotion of the dataset. Although the Turkers have experienced numerous samples of these emotions in their daily lives, they are still considered ``naive'' for this dataset. \par
While it is feasible to train computer systems for specific types of data (e.g., in a certain environment), it is often not possible to provide similar training to humans and hence they will always be operating in the ``naive'' mode.  Some applications where trained automatic classification systems can replace naive human coders include: 1) warning managers at call centers when either a customer or the customer service representative displays a negative emotion (such as anger, frustration, etc.); 2) applications where there is sensitive data and the content should not be released to human workers due to privacy issues; 3) a vehicular application that warns a driver about negative emotions to avoid road rage; and 4) applications that help those unable to decode emotions accurately, such as those with autism or certain cognitive degeneration diseases.  In these systems, it is not required to classify every ``sample'' (e.g., each 2-3 s of audio); instead, it is important that when an emotion classification label is added to the data, that classification is accurate.  As shown in our study, an automatic classification system is able to meet this requirement, providing quite high accuracy values by classifying between 50\% and 80\% of the data. \par
One interesting question for future work is how quickly humans would be able to be trained on a particular dataset, and once trained, would they be able to provide accuracy performance similar to the automatic classification system?  If humans could be trained quickly, then this would be a feasible option for some applications; however, if the cost (time and resources) to train humans is large, the automatic classification system remains an attractive alternative.
\newpage
\section{Conclusion}
This study compares the performance of a speech-based automatic emotion classification system with the performance of naive human coders in classifying emotions for speech utterances. The results show that the automatic system achieves much better accuracy in almost all cases.  Additionally, the automatic system can improve the classification accuracy by rejecting to classify samples for which it is not confident in the classification, while the naive human coders were not able to improve their accuracy through specifying their confidence in their classification.  These results show that a speech-based automatic emotion classification system is feasible as a replacement for applications that utilize naive human coders to classify emotion.
\balance
\bibliographystyle{abbrv}
\bibliography{refs}

\end{document}